\documentclass[a4paper,superscriptaddress,twocolumn,showkeys,pre,aps]{revtex4}

\usepackage[utf8]{inputenc}
\usepackage[english]{babel}
\usepackage{amssymb}
\usepackage{amsmath}
\usepackage{graphicx}
\usepackage[caption=false]{subfig}
\usepackage{float}

\usepackage[normalem]{ulem}
\usepackage{color}

\definecolor{myred}{rgb}{0.6,0,0}
\definecolor{myblue}{rgb}{0,0,0.6}
\definecolor{sigcol}{RGB}{17,97,165}
\definecolor{shadecolor}{RGB}{211,220,238}
\definecolor{edgecolor}{RGB}{17,97,165}

\newcommand{\Hl}[1]{\textit{#1}}

\renewcommand{\subsubsection}[1]{\textit{\textbf{#1.}}}

\newcommand{\halfgr}[1]{\includegraphics[width=1\linewidth]{#1}}

\renewcommand{\subsubsection}[1]{}
\newcommand{\sifig}[1]{\includegraphics[width=1.\linewidth]{#1}}

\begin{document}

\author{Kaj-Kolja Kleineberg}
\affiliation{Departament de F\'isica Fonamental, Universitat de Barcelona, Mart\'i i Franqu\`es 1, 08028 Barcelona, Spain}
\author{Mari\'an Bogu\~{n}\'a}
\affiliation{Departament de F\'isica Fonamental, Universitat de Barcelona, Mart\'i i Franqu\`es 1, 08028 Barcelona, Spain}
\date{\today}
\title{Evolution of the digital society reveals balance between viral and mass media influence}

\begin{abstract}
Online social networks (OSNs) enable researchers to study the social universe at a previously unattainable scale. The worldwide impact and the necessity to sustain their rapid growth emphasize the importance to unravel the laws governing their evolution. We present a quantitative two-parameter model which reproduces the entire topological evolution of a quasi-isolated OSN with unprecedented precision from the birth of the network. This allows us to precisely gauge the fundamental macroscopic and microscopic mechanisms involved.  Our findings suggest that the \hbox{coupling} between the real pre-existing underlying social structure, a viral spreading mechanism, and mass media influence govern the evolution of OSNs. The empirical validation of our model, on a macroscopic scale, reveals that virality is four to five times stronger than mass media influence and, on a microscopic scale, individuals have a higher subscription probability if invited by weaker social contacts, in agreement with the ``strength of weak ties'' paradigm.
\end{abstract}

\keywords{complex systems | complex networks |  evolution of online social networks}

\maketitle

\section{Introduction} 

The rapid growth of online social networks (OSNs), like Twitter or Facebook,
is reshaping the social landscape and changing the way humans interact on a world-wide scale. Needless to say, social networks existed well before OSNs were even invented. However, OSNs offer us the unprecedented opportunity to map social interactions at a scale that was unattainable before the digital era. This has transformed OSNs into huge sociological laboratories, boosting social sciences up to the level of experimental sciences. There is, however, an important difference between traditional social networks and OSNs. Technology-mediated social interactions constitute accelerating phenomena already observed in conventional social networks. Nevertheless, in the case of OSNs these take place faster and on a world-wide scale. This is already changing the way companies try to promote or sell their products  with viral marketing campaigns~\cite{Leskovec:viral_marketing,Richardson:viral_marketing,Moro:viral_dynamics,Moro:viral_marketing}, the way influential people, e. g. politicians, interact with their followers in Twitter~\cite{twitter:viral_diffusion,Twitter:virality,Grabowicz:2012fk}, or the way people self-organize and cooperate in protest movements~\cite{cooperation:pnas,arabspring:socialmedia}, crowdfunding~\cite{crowdfunding}, etc.

Nonetheless, this socio-technological revolution must come along with the development of new technologies able to sustain its growth. At this respect, one important issue concerns the design of the basic architecture of current OSNs. The best way to solve the scalability limitations of OSNs is to replace their centralized architecture by a fully decentralized one~\cite{isocial}. This can address privacy considerations and improve service scalability, performance, and fault-tolerance in the presence of an expanding base of users and applications. However, to accomplish this program successfully, it is necessary to take into account the social structure of the underlying society and how it interacts with the system, a task that involves network, computer, and social sciences.

As a matter of fact, we already have a fairly good knowledge on the topological properties of the ``social graph'' among users of OSNs~\cite{Mislove:2007fk,analysis_huge_osn,anatomy:facebook}. Indeed, large datasets of OSNs have allowed researchers to characterize their topology and to validate many principles from the social sciences, like the \Hl{``six degrees of separation''}~\cite{Milgram1967,Travers69anexperimental,newman:structure,albert1999dww,Kleinberg:smallworld} by S.~Milgram or the \Hl{``strength of weak ties''} by M.~S.~Granovetter~\cite{granovetter1973,Friedkin:weakties,Lu:role_of_weak_ties,Onnela:mobile,Thurner:multi,Ugander:2012}. However, these results concern static snapshots of the system and, thus, offer little insights into the fundamental mechanisms leading to the evolution of OSNs. Such insights can only be obtained from a detailed analysis of the temporal evolution of topologies of OSNs~\cite{Kumar2006,Hu2009,Hu2009a,Li,Aiello2010}. As we shall show, in the case of  real OSNs, such temporal evolution follows an intricate path: an initial phase where the social graph is made of small clusters with increasing diameter and average degree, followed by a dynamical percolation transition and, finally, an epoch of increasing average degree and shrinking diameter, akin to the observations by J.~Leskovec, J.~Kleinberg, and C.~Faloutsos in~\cite{Leskovec2007,Leskovec2005}. Interestingly, this type of history cannot be explained by standard models of growing networks under preferential attachment-like mechanisms, thus calling for new fundamental principles. 

In this paper, we focus on a particularly important case study, the Slovakian OSN ``Pokec''~\cite{Takac2012}. This network has a combination of unique properties that make it the perfect testbed for our purposes, namely,  the following: i) It is the most popular friendship-oriented OSN in the country. ii) Its size represents $25\%$ of the country's population. However, a simple demographic analysis of both the country and Pokec users suggest that, with its current size, it is covering a large fraction of the population susceptible to ever participate in OSNs. iii) The slovak language is mostly spoken within the country and iv) we can reconstruct its temporal evolution since its birth. As a result, we have the full history of a quasi-isolated OSN whose final state is also a good proxy for the underlying off-line friendship network. We hypothesize that such underlying social structure is essential for the emergence of OSNs. Under this premise, we introduce a simple model that incorporates viral dynamics and mass media influence operating on a multiplex network formed by the on- and off-line social graphs. The model reproduces very well the topological evolution of the Pokec OSN. Nevertheless, the perfect match is only achieved by introducing into the model the ``importance of weak ties'' paradigm, yet another empirical evidence in support of Granovetter's theory.

\section{Results}

\begin{figure}[t]
\centering
\halfgr{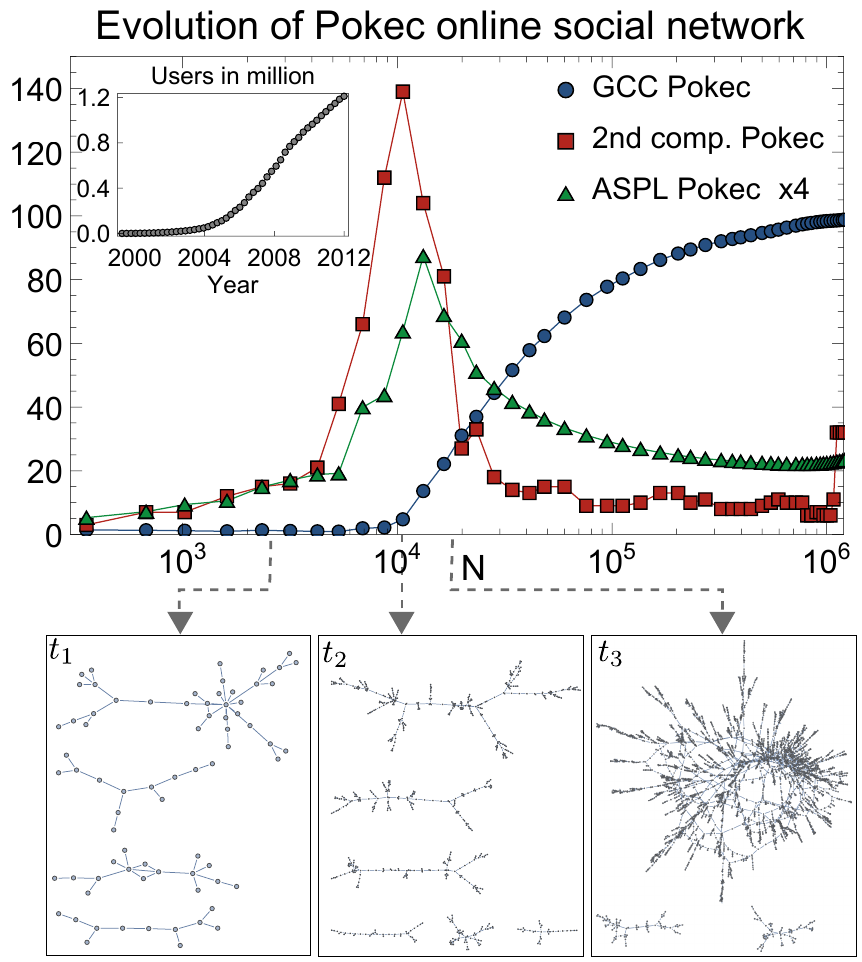}
\caption{Topological evolution of the empiric network. \textbf{Top:} The inset shows the evolution of the network size from 1999 to 2012. The main plot shows the relative size of the GCC (blue circles), the size of the second largest component (red squares), and the average shortest path length (green triangles, multiplied by four for better readability). \textbf{Bottom:} The largest components of the network are visualized at three different times, before the critical point, $t_1$, at the critical point, $t_2$, and after it, $t_3$. \label{fig_pokec_size}}
\end{figure}

\subsection{Evolution of the OSN Pokec: An example of a dynamical percolation transition}
\label{section_empiric_data}
   
\subsubsection{The dataset}
 
Pokec is by far the largest and most popular friendship-oriented OSN in Slovakia~\cite{Takac2012}. By April of 2012, it gathered around $1.6$ million users and $30$ million directed friendship relations. Nevertheless, not all directed links correspond to a real social tie: Alice might consider Bob as her friend while Bob may not have the same consideration for Alice. Thus, we discard all non-bidirectional links  from the original graph and treat those left as undirected edges. The resulting filtered network is composed of $1.2$ million users and 8.3 million bidirectional friendship connections. Interestingly, available users profile data contains the registration date of all users. Using this information, we can replay the history of the network topology by assuming that an edge between two users exists at a certain time if both users exist at that time. This approximation is reasonable due to observations from e.g.\cite{Li}, which suggests that most edges are created in a short time period after the birth of its end nodes. 
   
\subsubsection{Phase transition from disconnected to connected phase}
In the inset of Fig.~\ref{fig_pokec_size}, we show the temporal evolution of the number of registered users. We clearly appreciate a sustained monotonous increase, suggesting that the popularity of Pokec has not diminished even after the onset of Facebook in the year 2004. This monotonous relation allows us to use the number of current users, $N(t)$, as a measure of time instead of the physical time $t$. While this is only a rescaling of the temporal axis, it makes the comparison with models easier. The main plot in Fig.~\ref{fig_pokec_size} shows the evolution of the giant connected component (GCC) as a function of the network size. We observe a behavior that could be interpreted as a dynamical phase transition between a phase which consists of small disconnected clusters and a percolated phase where a macroscopic fraction of the network (99\% at the end of the evolution) is connected. In percolation theory, the signature of such continuous transition is encoded in the divergence, at the critical point, of the susceptibility $\chi$, defined as a measure of the ensemble fluctuations of the size of the GCC. Unfortunately, this technique cannot be applied in our case as the temporal evolution of the Pokec social graph is just one realization of the process. An alternative of the susceptibility is the size of the second largest connected component. This measure is known to diverge at the critical point and, in a single realization of a finite system, it shows a maximum close to the critical percolation point. 

Figure~\ref{fig_pokec_size} shows a clear peak in the size of the second largest connected component, indicating that, indeed, we are observing a dynamical phase transition~\footnote{The distribution of sizes of disconnected components at this point is a power law, another clear indication of the presence of a continuous phase transition, see appendix.}. In the same figure, we also show the behavior of the average shortest path length (ASPL) within the largest connected component, which shows a quite interesting behavior. During the first stage of the evolution, the ASPL increases with the network size but its growth is not compatible with a logarithmic law, as predicted by the small-world effect. Shortly after the critical point, the ASPL reaches its maximum and then decreases while the size and the average degree (see Fig.~\ref{fig_local_meandegree}~B) of the network increase. At long times, the ASPL reaches a value of about five, which is compatible with the small world effect. This behavior was first observed in~\cite{Leskovec2007}.
   
  \begin{figure}[t]
  \centering
   \halfgr{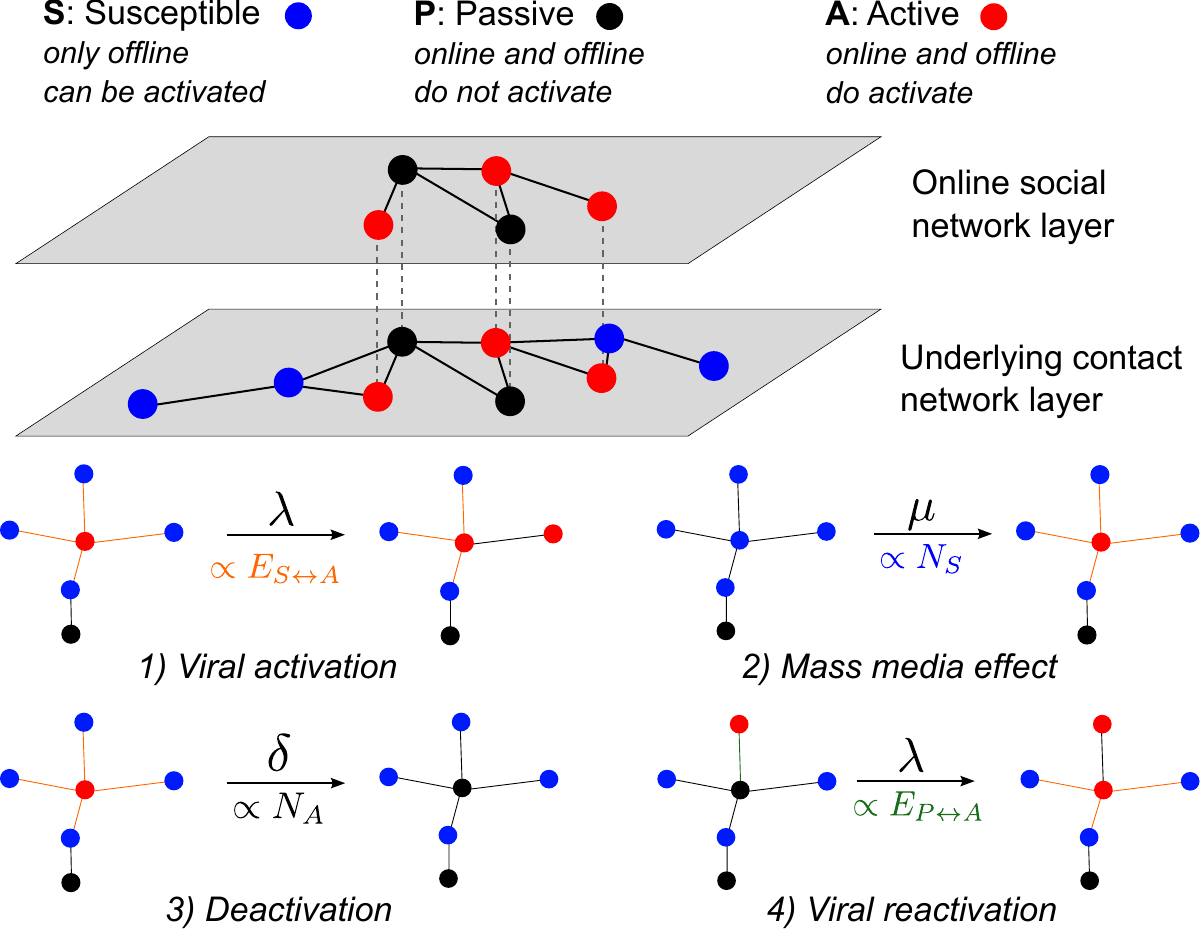}
    \caption{Illustration of the two-layer model. The upper layer represents the online social network and contains the active and passive nodes. The bottom layer corresponds to the underlying contact network which contains all nodes. The four dynamical processes are shown in topological illustrations below.   \label{fig_multi_sketch}}
   \end{figure}
   
   \subsection{Basic model: trade-off between virality and mass media influence}
\label{section_model}
 \subsubsection{Model mechanism}

This anomalous behavior challenges current models for growing networks that are based on preferential attachment or similar mechanisms as, in general, they do not show dynamical percolation transitions. In this type of models, the pool of new nodes that are added to the system does not have any previous relation with existing nodes and the connections of newborn nodes are decided exclusively as a function of the current state of the network. However, in the case of OSNs, there is a pre-existing underlying off-line social network conditioning the growth of the OSN. Following this line of reasoning, we conjecture that the observed evolution is the result of a dynamical process that triggers potential users from the off-line social network to subscribe to the OSN. Under this assumption, nearly all dynamics able to induce the recruitment of all potential users will yield a dynamical percolation transition. Yet, different dynamics induce different temporal orders in the evolution of OSNs and, therefore, different topological histories.

Following these ideas, we design a two-layer multiplex model for the evolution of OSNs. The upper layer represents the online social network whereas the bottom layer represents the off-line social network. The latter can be considered the subgraph of all {\it a priori} susceptible individuals from the aggregation of all social interactions between individuals. Each individual can be in three different states depending on whether they are or are not enrolled in the OSN. Susceptible individuals are those not in the OSN but that might eventually become members of it. Active individuals belong to the OSN and are actively using it for their social interactions. Passive individuals also belong to the OSN but are not currently using it to interact with their social contacts, see Fig.~\ref{fig_multi_sketch}. The populations of susceptible, active, and passive individuals are governed by a combination of an epidemic-like process between active users and susceptible or inactive ones and a mass media effect which equally affects the population of susceptible individuals. There are four possible events.
\begin{enumerate}
 \item \Hl{Viral activation:} a susceptible node can be virally activated and added to the OSN by contact to an active neighbor in the traditional off-line network. This event happens at rate $\lambda$ per each active link.
 \item \Hl{Mass media effect:} each susceptible individual becomes active spontaneously at rate $\mu$ and is added to the OSN layer as a response to the visibility of the OSN.
 \item \Hl{Deactivation:} active users become spontaneously passive at rate $\delta$ and no longer trigger viral activations nor reactivate other passive nodes.
 \item \Hl{Viral reactivation:} at rate $\lambda$ an active user can reactivate a passive neighbor. The neighbor then becomes active and can trigger both viral activations and viral reactivations. 
\end{enumerate}
We can arbitrarily set $\delta=1$, which defines the timescale in units of the deactivation time. The model is then left with two independent parameters, the virality parameter $\lambda$ and the mass media parameter $\mu$. Finally, newborn users explore the OSN and connect to all their neighbors in the traditional off-line social network that, at the time of the subscription, are either active or passive. It is worth to point out that the dynamics between active and passive users is equivalent to the susceptible-infected-susceptible (SIS) epidemic model~\cite{anderson92}. As it happens in the SIS model, our model also has a critical rate $\lambda_c$ below which the number of active users vanishes whereas above it the activity of the OSN is self-sustained. This makes the model extremely versatile as it can explain the different fates of OSNs. We also note that a mean field version of this dynamics has been recently and independently proposed to model users' activity of OSNs~\cite{Bruno:Ribeiro}.

The viral activation and the mass media effect play complementary roles in terms of their impact on the topological growth of the network. The mass media effect is very likely to create new components especially at the beginning of the network evolution whereas the viral activation leads to the growth of already existing components. The interplay between these complementary principles is the fingerprint of the evolution of the online social network and the trade-off between these mechanisms governs the appearance of the phase transition. We will quantify and discuss this trade-off later.

Unfortunately, the rigorous validation of the model requires the precise knowledge of the topology of the underlying social network. However, in the particular case of the Pokec network, its large coverage among the subgraph of potential users suggests that we can consider the final snapshot of the Pokec OSN as a good proxy for the real underlying social network. Following this approach, we perform extensive numerical simulations of our model and compare the resulting evolution with the one we observe in the Pokec OSN. Of course, the real evolution of the Pokec network is still ongoing and, thus, we expect this approximation to fail as we approach the final size of the network. In particular, we do not expect the model to reproduce the network growth in physical time because as the model approaches the size of the empirical network a saturation process aparently slows down the dynamics. We deal with this problem by using the network size instead of physical time as the measure of the course of the evolution, which allows us to compare the topology of the model and the empiric network despite its ongoing evolution~\footnote{This corresponds to dynamically rescaling $\lambda$, $\mu$ and $\delta$ in the same way at each timestep. Hence the transformation preserves the fraction $\lambda/\mu$ which we discuss in terms of the trade-off between virality and mass media influence.}. 
\begin{figure}[t]
\centering
\halfgr{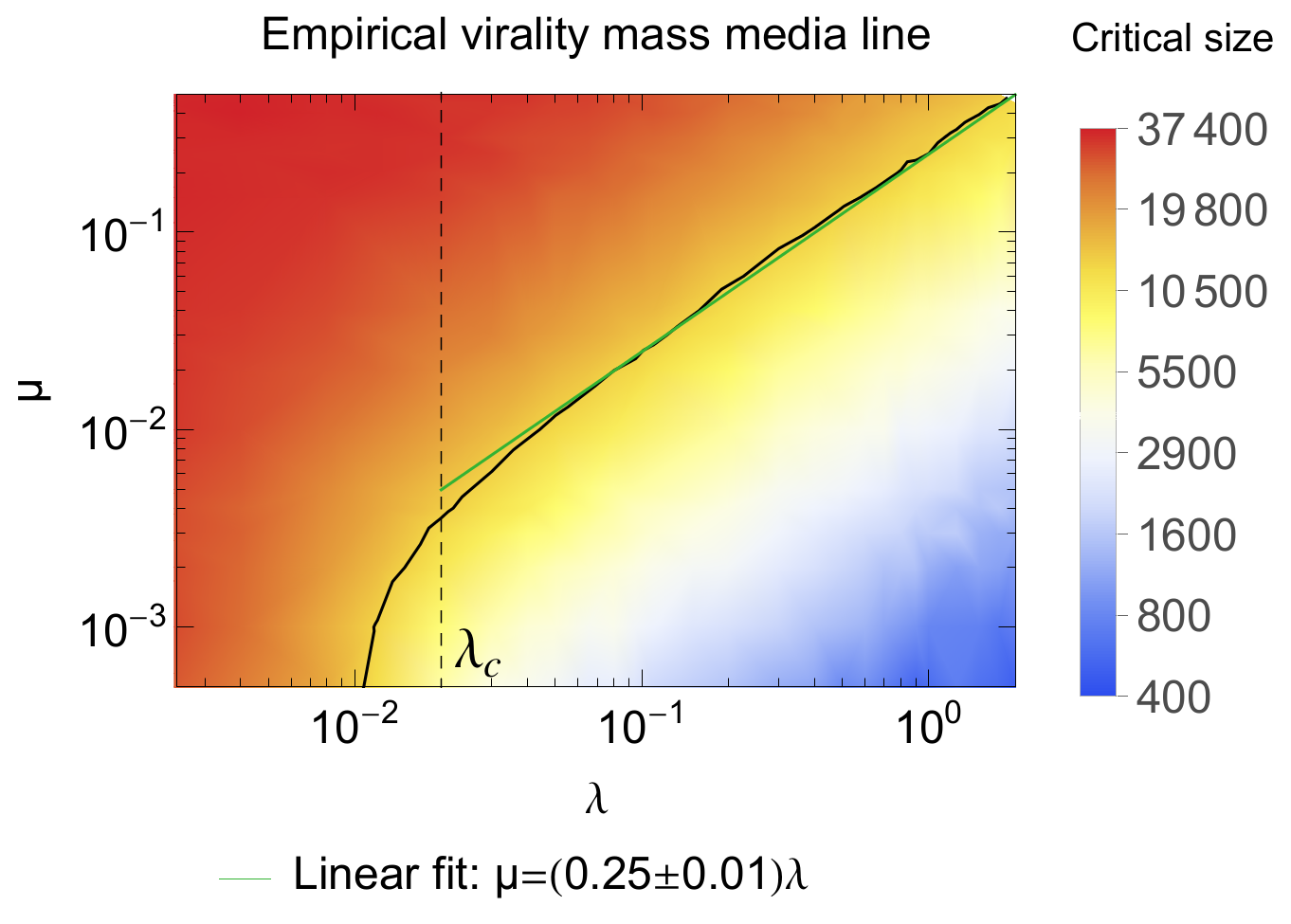} 
\vspace*{-0.7cm}
\caption{The color density plot represents the network size at the critical point for the respective parameters $\lambda$ and $\mu$. The solid black line indicates the virality mass media line corresponding to the critical size of Pokec ($N_c^P = 10600$). The green line shows a linear fit in the region above the sustained activity threshold $\lambda_c = 0.02$ (see appendix).
  \label{fig_density_plot}}
 \end{figure} 
   
\begin{figure}[t]
\centering
\halfgr{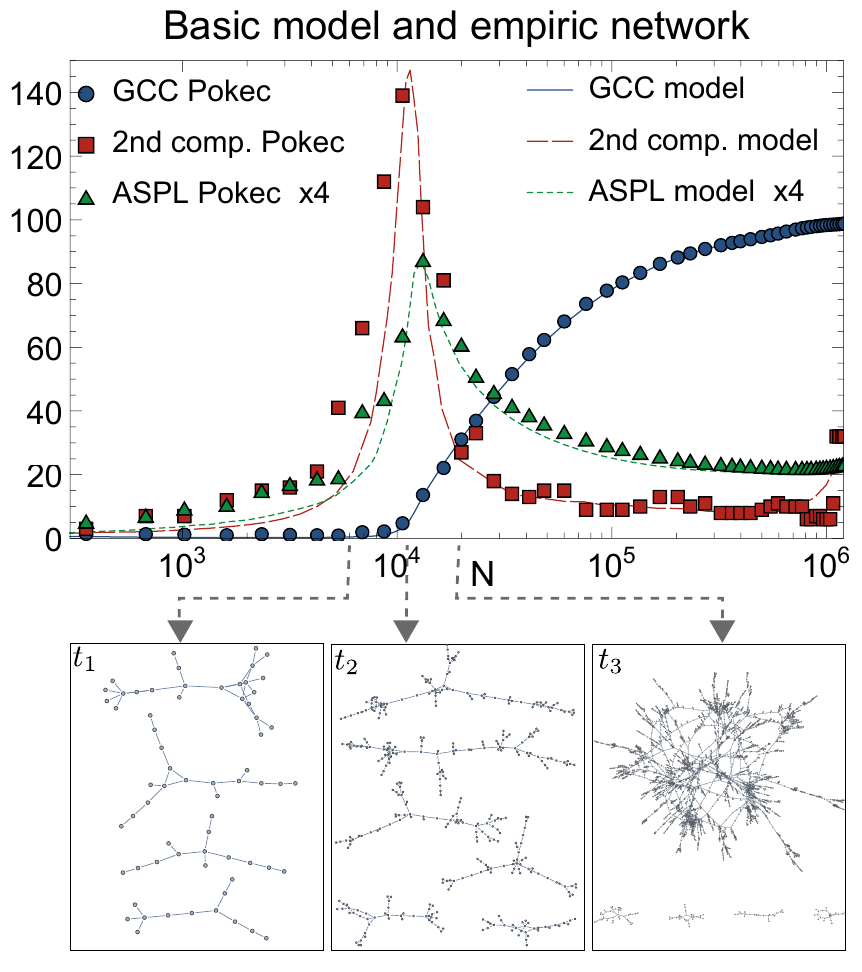}
\caption{Comparison of model and Pokec network evolution. \textbf{Top:} The symbols represent the empirical data, whereas the solid lines correspond to the results from the model averaged over 100 realizations with $\lambda = 0.03$ and $\mu = 0.008$. Points correspond to the empiric network and lines represent the results from the model. \textbf{Bottom:} Snapshots of the topology of the model at different times similar to Fig.~\ref{fig_pokec_size}. \label{fig_model_compare_transition}}
\end{figure}
   
 \subsubsection{Quantifying the trade-off between virality and mass media influence}

The results of our model show the emergence of a dynamical phase transition from a disconnected to a connected state. We take advantage of the uniqueness of the critical point to adjust the parameters of the model by matching the network size of the model and the empiric network at the transition point. To this end, we compute the critical size for different values of the parameters $\lambda$ and $\mu$, as shown in Fig.~\ref{fig_density_plot}. In the empiric network, the phase transition occurs at $N_c^{P} = 10600$ which is represented by the black contour line in the plot. The green line shows a linear fit according to 
\begin{equation}
\mu(\lambda) = (0.25\pm 0.01)\lambda \,.
\label{virality-mass media-eqn}
\end{equation}
The virality mass media line given by Eq.~\eqref{virality-mass media-eqn} quantifies the trade-off between the importance of the viral effect and the mass media effect for the evolution of the network. At the light of this result, we conclude that the viral effect is about four times stronger than the mass media effect. In other words, in the particular case of the Pokec OSN, it is four times more likely to subscribe to the network as a result of a friend's invitation than as the result of the information about the network available through the mass media. However, Eq.~\eqref{virality-mass media-eqn} only holds above a critical value of the virality parameter $\lambda>\lambda_c$, which corresponds to the critical threshold for the self-sustained activity of the network (see appendix). Below this limit, the virality mass media line bends downwards and, in the limit of $\lambda=0$, it is not possible to match the position of the critical size (see appendix). This implies that both virality and mass media influence are necessary and complementary mechanisms to explain the topological evolution of OSNs.

 \subsubsection{Phase transition and comparison with data}
\begin{figure}[t]
\centering
\halfgr{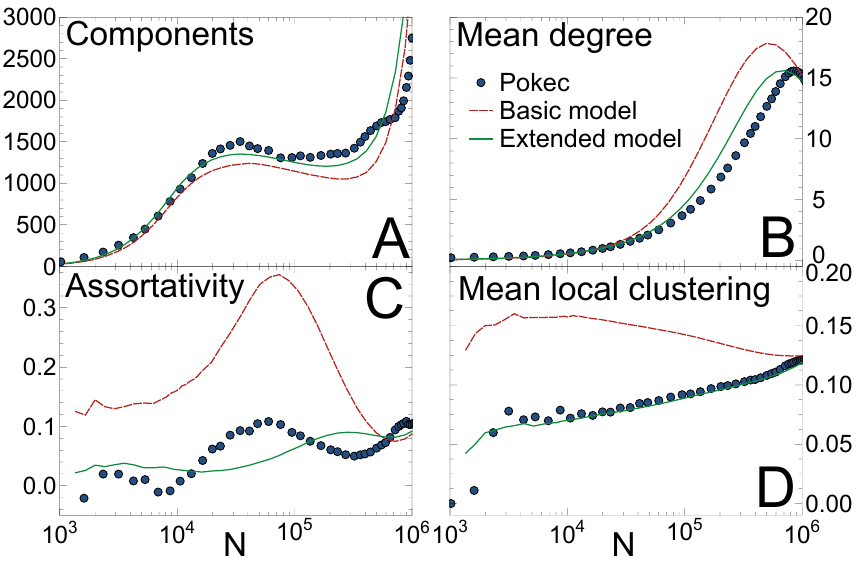}
\caption{Topological evolution of the empiric network (blue circles), the basic model (red dashed lines), and the extended model (green lines). \textbf{A:}~Evolution of the number of components of size $s > 1$. \textbf{B:}~The evolution of the mean degree shows a densification of the network. \textbf{C:}~The assortativity coefficient as defined in~\cite{Newman:introduction}. \textbf{D:}~The evolution of the mean local clustering coefficient (of nodes with $k>1$) exhibits an essential difference between the basic model (red dashed line) and the extended model with $\eta = -0.65$ (green line). 
\label{fig_compare_local} 
\label{fig_local_clustering}
\label{fig_local_assortativity}
\label{fig_local_meandegree}
\label{fig_compare_components}}
\end{figure}
In the active phase, the effect of changing the value of $\lambda$ is very mild if the relation Eq.~\eqref{virality-mass media-eqn} is preserved. In our case, we choose the value of $\lambda$ that best reproduces the evolution of the number of disconnected components (Fig.~\ref{fig_compare_components}~A) and obtain the corresponding $\mu$ from Eq.~\eqref{virality-mass media-eqn}. We then compare the results of the model with the empirical evolution of the Pokec OSN in Fig.~\ref{fig_model_compare_transition}. Interestingly, our two parameters model is able to reproduce the entire evolution of the network with an impressive precision for all measured global topological properties, such as the size of the giant component, the ASPL, and the size of the second largest connected component, see Fig.~\ref{fig_model_compare_transition}. However, the model is not able to reproduce the temporal trends of local quantities like the mean local clustering and assortativity coefficients, as shown in Figs. \ref{fig_compare_local} C and D. The clustering coefficient of the Pokec OSN steadily increases since the beginning of the evolution whereas the model exhibits first a sudden increase followed by a decreasing clustering coefficient. The assortativity coefficient fluctuates both in the model and in the empiric network, although its value in the model is about three times higher. This disagreement suggests that a local mechanism must be incorporated to reproduce simultaneously the global and local evolution of the network topology. In the next section, we present an extended version of our model which takes into account the overlap of each node's neighborhood, with interesting implications concerning the \Hl{``strength of weak ties''} paradigm.

\begin{figure}[t]
\centering
\halfgr{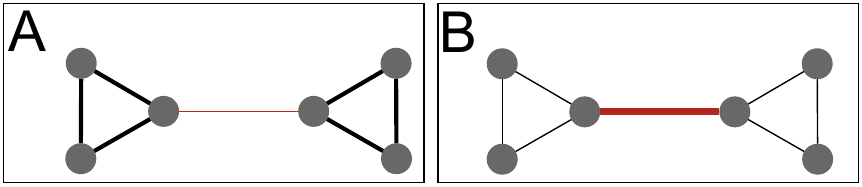}
\caption{\textbf{A:} Illustration of the strength of social ties defined by Eq.~\eqref{eqn_sij}. \textbf{B:} Illustration of viral transmissibilities for negative $\eta$ defined by Eq.~\eqref{eqn_weights}.
\label{fig_mult_ties}}
\end{figure}

\subsection{Extended model: strong vs. weak ties}
\label{section_weights}
 
  \subsubsection{Overlapping neighborhood and strength of social ties}

The viral activation mechanism of our model is completely blind to the network topology, that is, active users try to ``infect'' all their neighbors with the same probability. As a consequence, the model performs well at reproducing the evolution of the global topological quantities but it fails at reproducing trends in local quantities, like the clustering coefficient. However, according to Granovetter~\cite{granovetter1973}, the diffusion of information through a social tie is different depending on whether the tie is ``strong'' or ``weak''. Following Granovetter's idea, we use the overlap of two individuals' friendship network as a measure of the strength of their tie~\cite{granovetter1973}. In particular, given an edge connecting users $i$ and $j$, we define its social strength as 
\begin{equation}
 s_{ij} \equiv (m_{ij}+1) ,
 \label{eqn_sij}
\end{equation}
where $m_{ij}$ counts the number of triangles going through the edge or, equivalently, the number of common neighbors of the two users. 

Our previous model can now be easily extended to account for the strength of social ties. We assume that viral activation and reactivation through the edge $i \leftrightarrow j$ is given by
\begin{equation}
\lambda_{ij}=\lambda \frac{s_{ij}^{\eta}}{\langle s^{\eta} \rangle}.
\label{eqn_weights}
\end{equation}
The transmissibility-strength coefficient $\eta$ represents the relationship between the viral transmissibility and the strength of the social tie. When $\eta > 0$ viral transmissibility is proportionally to the strength of social ties, which puts special emphasis on the strong ties for the viral spreading. Instead, when $\eta < 0$ the highest viral transmissibilities are assigned to edges with low multiplicities, which tend to act like connectors between different clustered groups (see Fig.~\ref{fig_mult_ties}). In the case of $\eta = 0$, we have $\lambda_{ij}=\lambda$ and we recover the basic model discussed in the previous section.

\subsubsection{Quantifying the transmissibility-strength relationship}
  \begin{figure}[t]
   \centering
     \halfgr{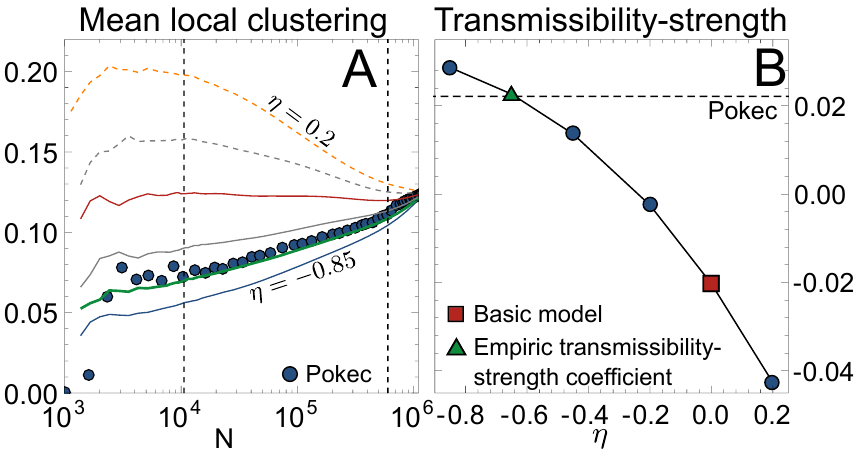}
    \caption{\textbf{A:} Evolution of the mean local clustering coefficient for different values of $\eta$ from $-0.85$ to $0.2$. The lines correspond to model results for different $\eta$ and the symbols represent the evolution from the empiric network. We set $\lambda = 0.03$ and $\mu = 0.006$. \textbf{B:} Slopes of the evolution of the mean local clustering coefficient for different values of the transmissibility-strength coefficient. The dashed line represents the slope of the clustering evolution of the Pokec network. The empirical transmissibility-strength coefficient $\eta = -0.65$ is given by the intersection of this line with the curve representing the model results.
    \label{fig_eta}}
   \end{figure}

We quantify the transmissibility-strength coefficient $\eta$ by comparing the evolution of the mean local clustering coefficient in the Pokec OSN with results from the extended model. Figure~\ref{fig_local_clustering}~D shows that the clustering coefficient of the Pokec OSN grows approximately linearly with the logarithm of the network size. Thus, we interpolate the evolution of the clustering coefficient of our model for different values of $\eta$ and compare the obtained slopes with the empirical one, as shown in Fig.~\ref{fig_eta}. The extended model exhibits an increasing clustering coefficient for $\eta < -0.2$ (see figure~\ref{fig_local_clustering}~D) and the best match with the Pokec OSN is achieved at the value of $\eta = -0.65$ which is, remarkably, a negative value. This is yet another empirical proof of Granovetter's theory on the importance of weak ties in processes of diffusion of information in social networks~\cite{granovetter1973}. An alternative empirical validation of the same principle was provided in~\cite{Ugander:2012}, where it was found that the probability of accepting an invitation to join an OSN is not proportional to the number of social contacts of the invited individual but to the number of different social contexts --the structural diversity-- within the individual's life. Notice that a similar effect is achieved in our model when the exponent $\eta$ is negative.

The introduction of weighted transmissibilities in our model does not affect significantly the evolution of the global topological properties. Indeed, for $\eta=-0.65$, the virality mass media line behaves like in the basic model with the difference that now the relation between $\lambda$ and $\mu$ is 
\begin{equation}
  \mu(\lambda) \approx (0.21\pm0.01)\lambda . 
  \label{eqn_line_ext}
  \end{equation}
As for the rest of topological measures, the evolution of the GCC, the size of the second largest connected component, the ASPL, and the diameter are basically identical to the case of the basic model and are shown in the appendix. In Fig.~\ref{fig_compare_local}, we show results for the number of components, the mean degree, the assortativity coefficient, and the mean local clustering, which are all in very good agreement with their empirical counterparts.

\section{Discussion}

Comprehensive datasets on the evolution of OSNs offer us the opportunity to determine the principal mechanisms involved in social contagion and online activity of individuals. At this respect, the OSN Pokec, with its peculiar evolution and being almost isolated, is particularly appropriate. We have shown that the evolution of Pokec's topology can be explained very precisely on a quantitative level by a two-layer model, which accounts for the underlying real social structure, combined with two main mechanisms. First, a viral effect, responsible for the social contagion of new users and, second, a mass media effect, leading to random subscriptions of new users. Interestingly, the trade-off between these two mechanisms is what governs the topological growth of OSNs. In the particular case of the OSN Pokec, the quantification of this trade-off tells us that the viral effect is between four to five times stronger than the mass media effect. This can explain the proliferation of viral marketing campaigns, in detriment of traditional advertising~\cite{Leskovec:viral_marketing}. To our knowledge for the first time a model with only very few parameters yields quantitatively precise insights about the topological formation of OSNs. This makes our model a necessary foundation for the development of next generation online social networking services. 

Beyond the global behavior of our basic model, the social neighborhood of individuals has shown to be crucial to explain the evolution of local topological quantities in Pokec. We find that viral transmissibility is inversely proportional to the strength of social ties. This result is particularly interesting as it corroborates recent empirical findings concerning the role of ``structural diversity'' on social contagion processes by analyzing email invitations from Facebook users~\cite{Ugander:2012}. However, our model allows us to identify and quantify this effect exclusively from --and hence its impact on-- the topological evolution of the OSN. Alongside with Granovetter's conclusion about the importance of weak ties for individual success, our results give rise to the interpretation that OSNs evolve in a way to improve the possibilities for individual success. This might constitute an important reason for the huge popularity of OSNs.

Our findings here suggest interesting future research lines. Indeed, the particular OSN analyzed in this paper is a quasi-isolated system and, thus, allows us to gauge the fundamental mechanisms at play in the evolution of OSNs. However, in a general situation, an entire ecosystem of OSNs operate simultaneously, competing for the same users, which now become a scarce resource. The introduction of competition among OSNs in our model opens the possibility to develop an ecological theory of the digital world. 

\begin{acknowledgments}
We thank Pol Colomer de Sim\'on and Guillermo Garc\'ia P\'erez for useful comments and suggestions, Jure Leskovec for providing the Stanford Large Network Dataset Collection as well as Lubos Takac and Michal Zabovsky for the Pokec dataset~\cite{Takac2012}.
This work was supported by the European Commission within the Marie Curie ITN ``iSocial'' grant No.\ PITN-GA-2012-316808; by a James S. McDonnell Foundation Scholar Award in Complex Systems; by the ICREA Academia prize, funded by the {\it Generalitat de Catalunya}; by MICINN project No.\ FIS2010-21781-C02-02; by {\it Generalitat de Catalunya} grant No.\ 2009SGR838; M.~B. acknowledges support from the European Commission LASAGNE project No.\ 318132 (STREP).
\end{acknowledgments}


\appendix

\section{Null model}
 \label{subsection_nullmodel}
    
       \begin{figure}[H]
   \centering
     \sifig{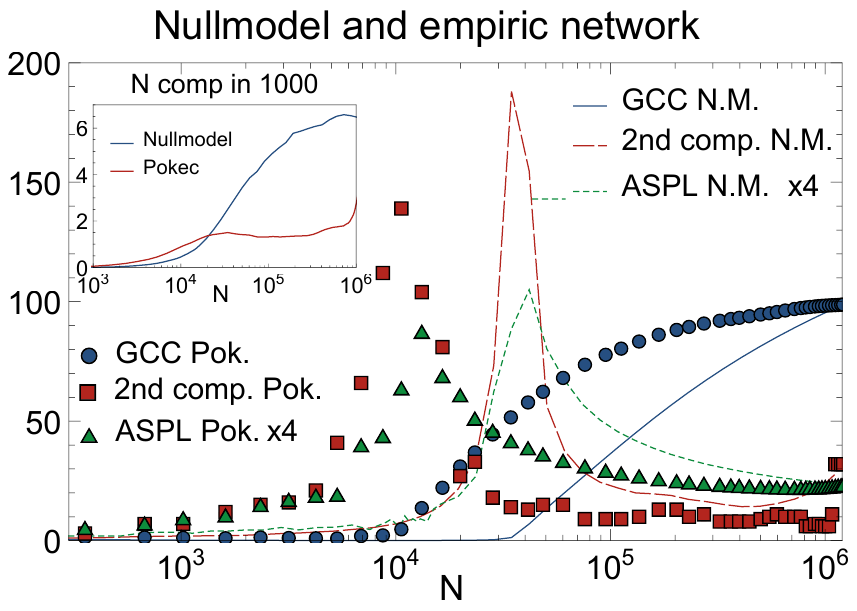}
    \caption{A null model with underlying Pokec network which consists of 
randomly adding nodes to the online social network. Points correspond to the 
empiric network and solid lines correspond to the null model.  \label{fig_null}}
   \end{figure}
    
     In figure \ref{fig_null} we present the results from a null model for the 
role of the underlying empiric network. We again use the Pokec network as 
underlying and add nodes completely randomly to the network. Note, that this 
corresponds to our model for $\lambda = 0$ and arbitrary $\mu > 0$. However, the 
choice of $\mu$ then just fixes the model timescale, which we adjust implicitly 
by transforming physical time to the intrinsic network timescale given by the 
number of nodes. We observe, that the phase transition takes place at a larger 
network size. Note, that there is no more parameter to adjust. The number of 
components (see inset in fig. \ref{fig_null}) also varies strongly between the 
empiric network and the presented null model. 
          
   We conclude, that the occurrence of the phase transition is included in the 
structure of the underlying network. Nevertheless, a null model with exclusively 
random subscriptions fails to reproduce the critical point of the phase 
transition as well as the evolution of the number of components.

   \section{Sustained activity threshold}

   \begin{figure}[H]
       \centering
    \sifig{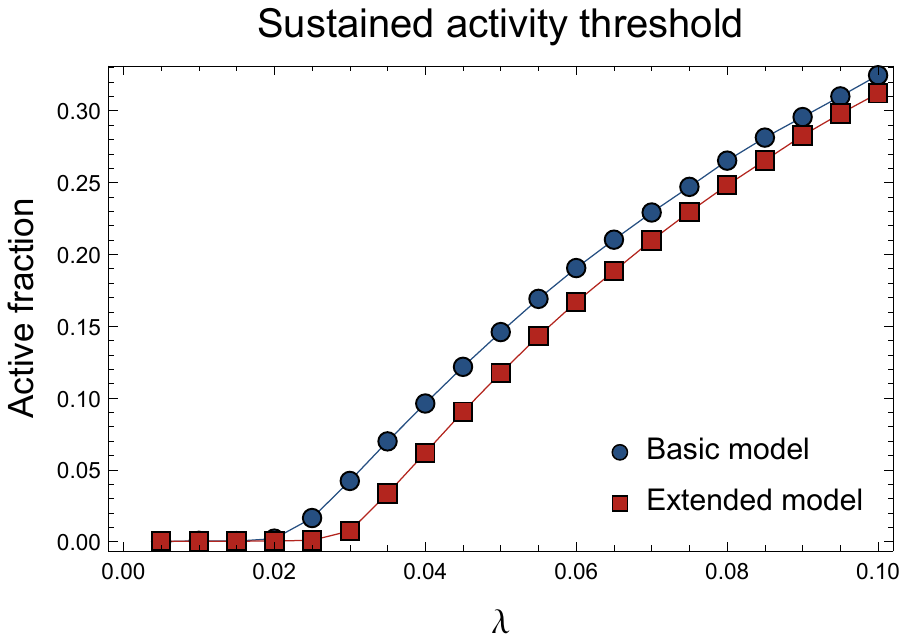}
    \caption{Sustained activity threshold $\lambda_c \approx 0.02$. Below this 
threshold, the activity of the network is not sustained and eventually the whole 
network will become passive. \label{fig_lambda_crit}}
   \end{figure}

From figure \ref{fig_lambda_crit} we obtain the critical parameter $\lambda_c$ 
for the outbreak of the SIS equivalent PAP dynamics within the online social 
network layer. We obtain
\begin{equation}
 \lambda_c \approx 0.02
\end{equation}
which appears to be quite robust to the assignment of weights. Below this 
threshold, the whole network become passive. We suggest, that this corresponds 
to the practical disappearance of the network as observed in many empiric online 
social networking services.

   \section{Pathlength and diameter}
   
    \begin{figure}[H]
    \centering
    \includegraphics[width=1\linewidth]{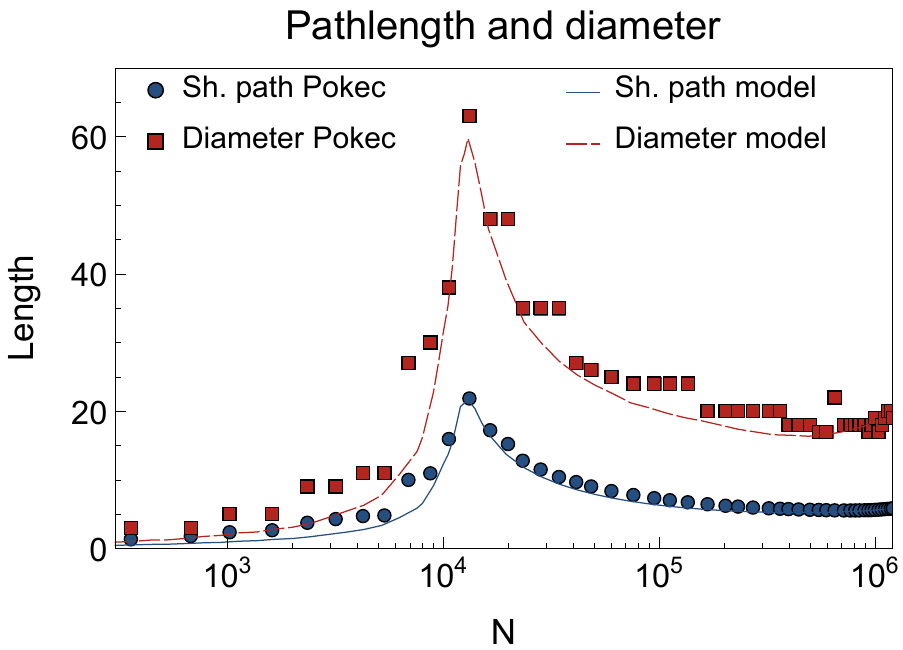}
         \includegraphics[width=1\linewidth]{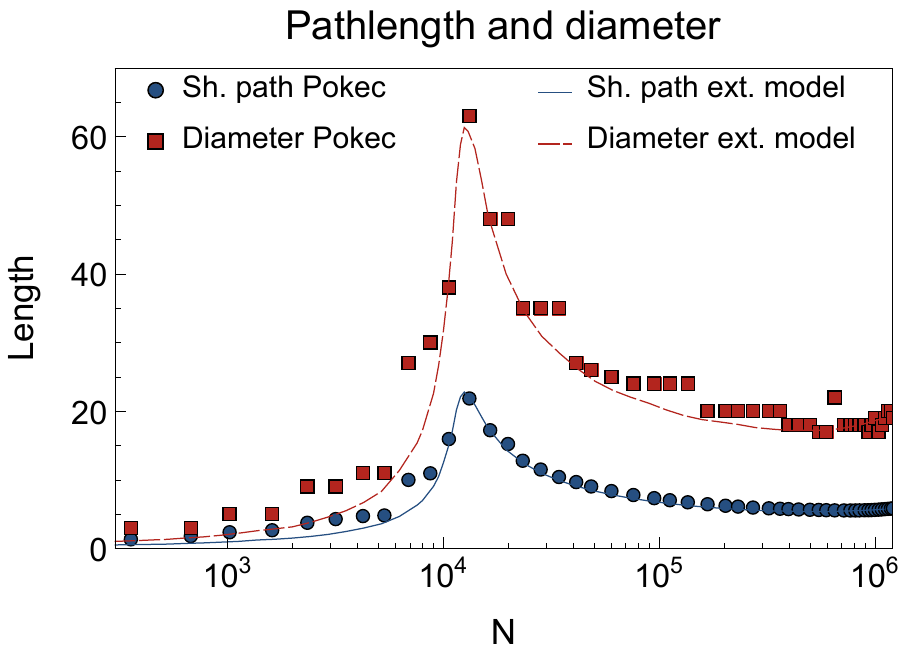}
    \caption{Pathlength and diameter for basic model (top) and extended model 
(bottom).   \label{fig_path_dia} \label{fig_model_extended}}  
   \end{figure}
   
   We observe the same behavior in the evolution of the average shortest path 
length and the network diameter (see figure \ref{fig_path_dia}). In the 
connected regime the pathlength and the diameter within the GCC decrease.
   In the disconnected regime, the average shortest path length and the diameter 
increase and reach their maximum at the critical point. The behavior of the 
extended model is equivalent.
   
\section{Explicit results for extended model}

\begin{figure}[H]
    \centering
\sifig{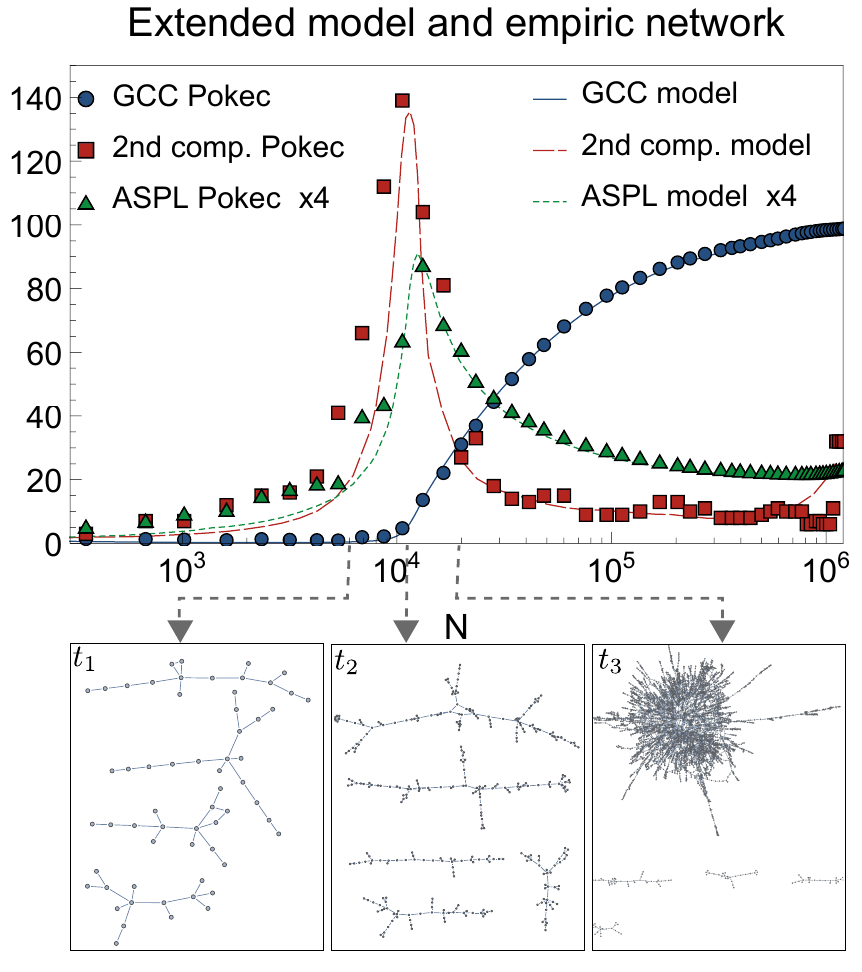}
 \caption{GCC (blue), size of second largest component (red), and average 
shortest path length (green) for the extended model for the parameters $\eta = 
-0.65$, $\lambda = 0.03$, and $\mu = 0.006$. \label{fig_ext}}
\end{figure}
\begin{figure}[H]
    \centering
\sifig{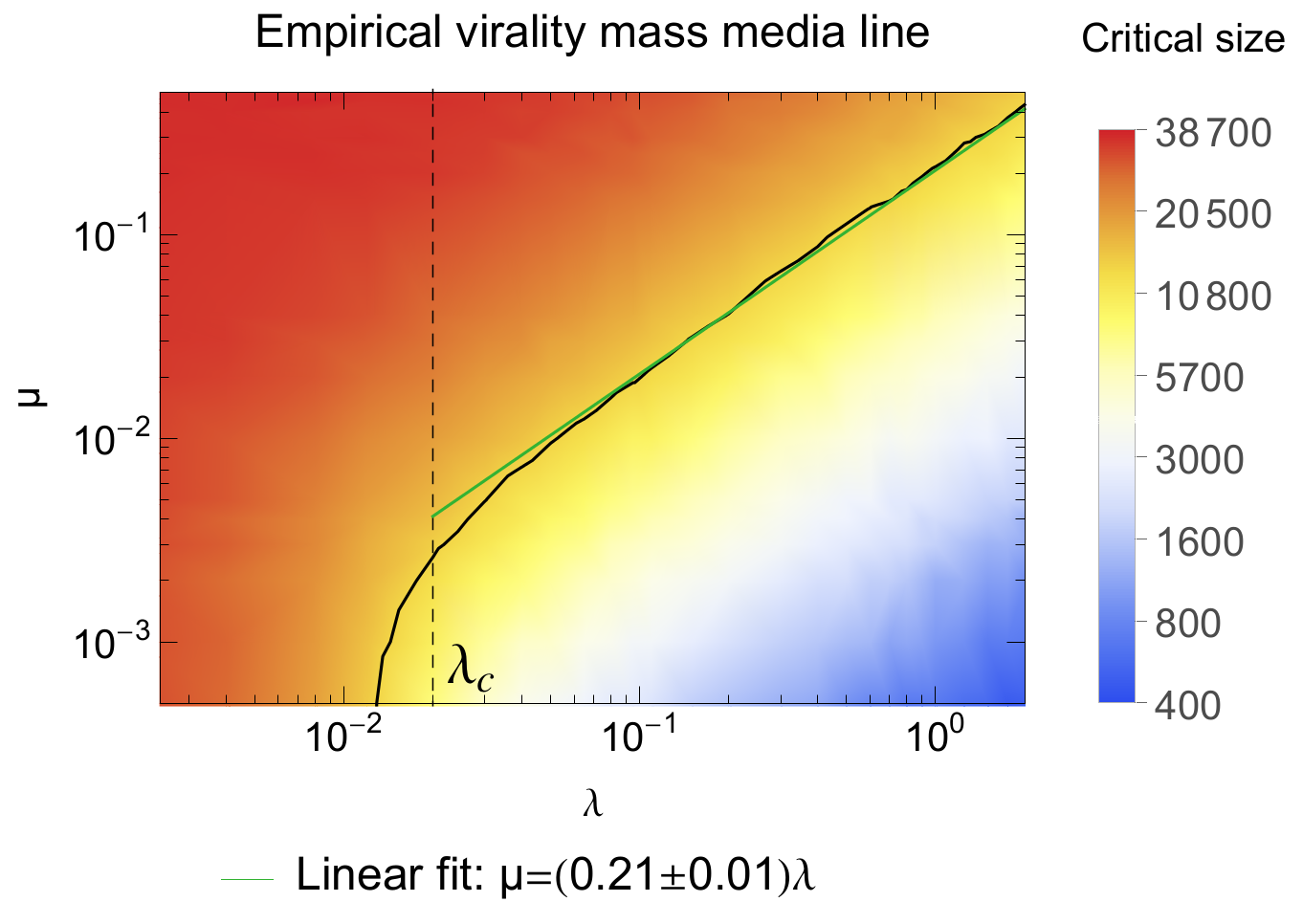}
 \caption{Matching of the critical point for the extended model. 
\label{fig_ext_line}}
\end{figure}

\section{Distribution of component sizes}
\begin{figure*}[t]
    \centering
 \includegraphics[width=0.32\linewidth]{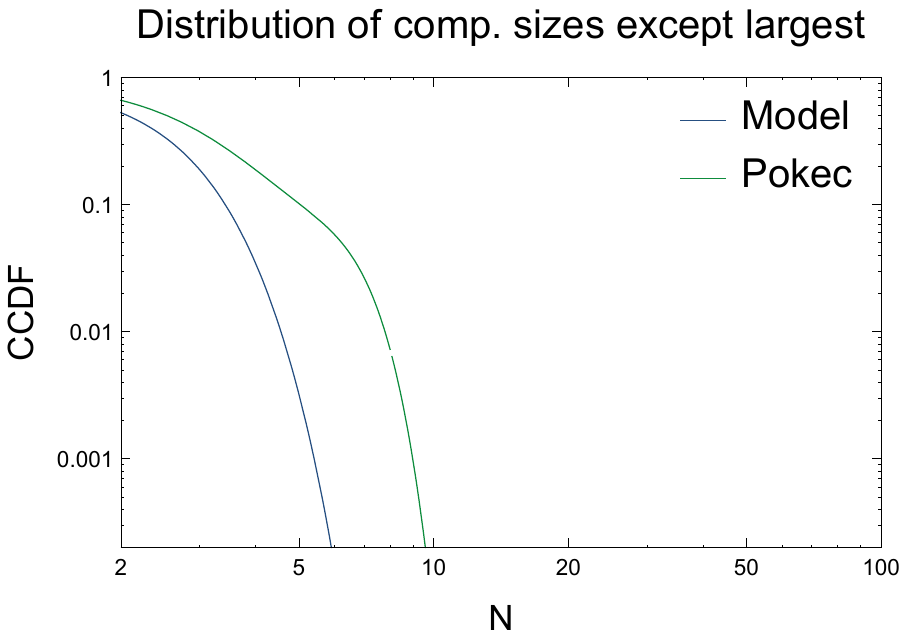}
  \includegraphics[width=0.32\linewidth]{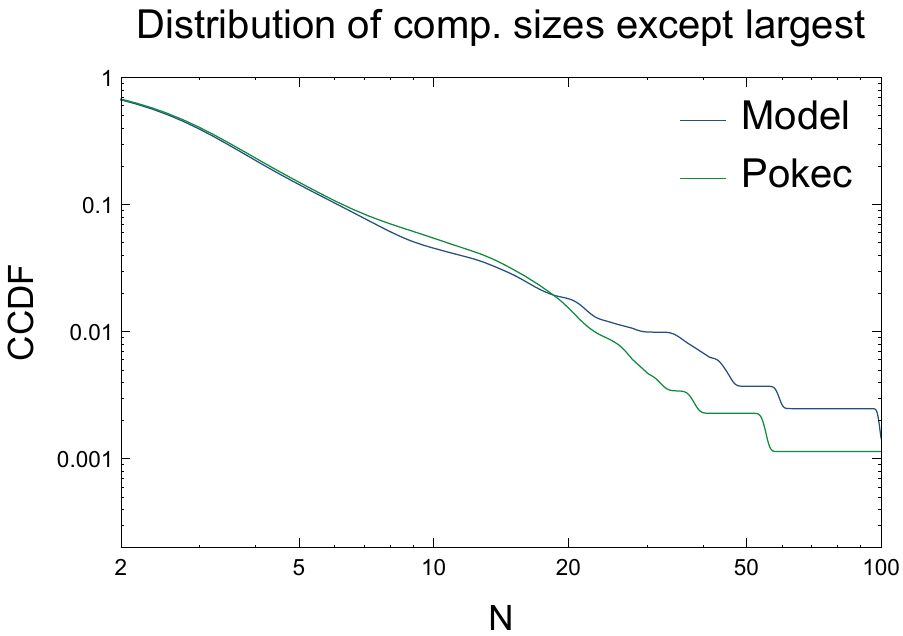}
   \includegraphics[width=0.32\linewidth]{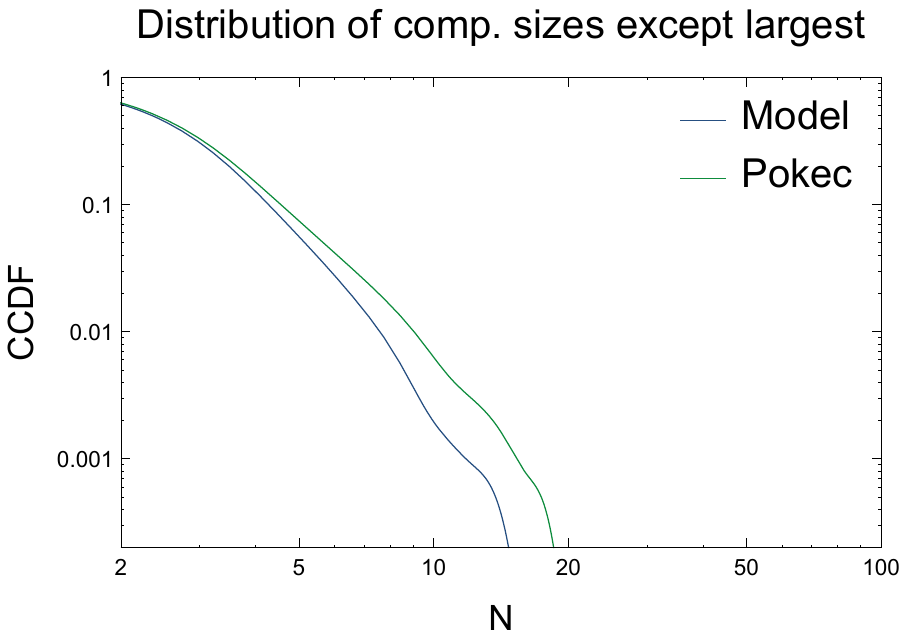} \\
    \includegraphics[width=0.32\linewidth]{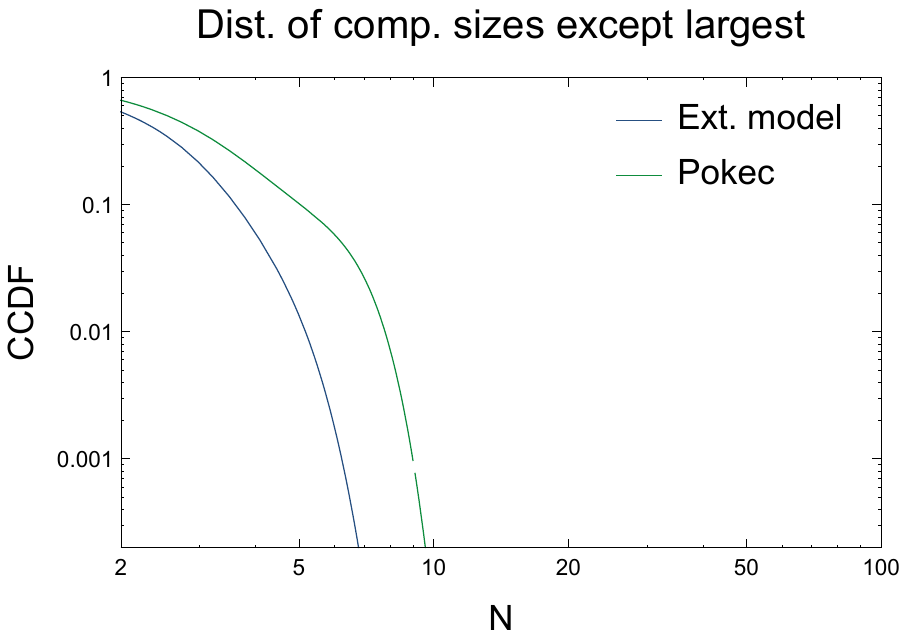}
  \includegraphics[width=0.32\linewidth]{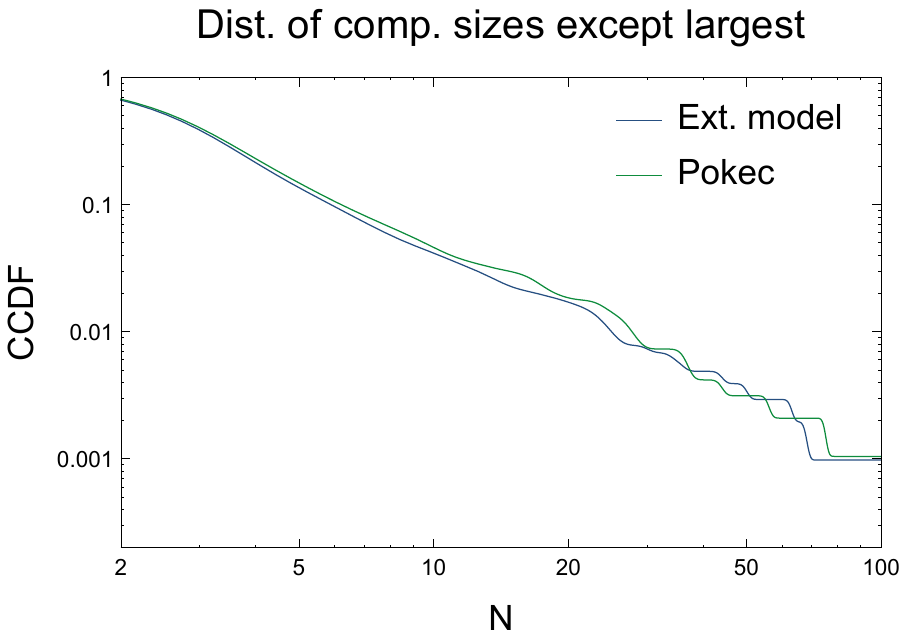}
   \includegraphics[width=0.32\linewidth]{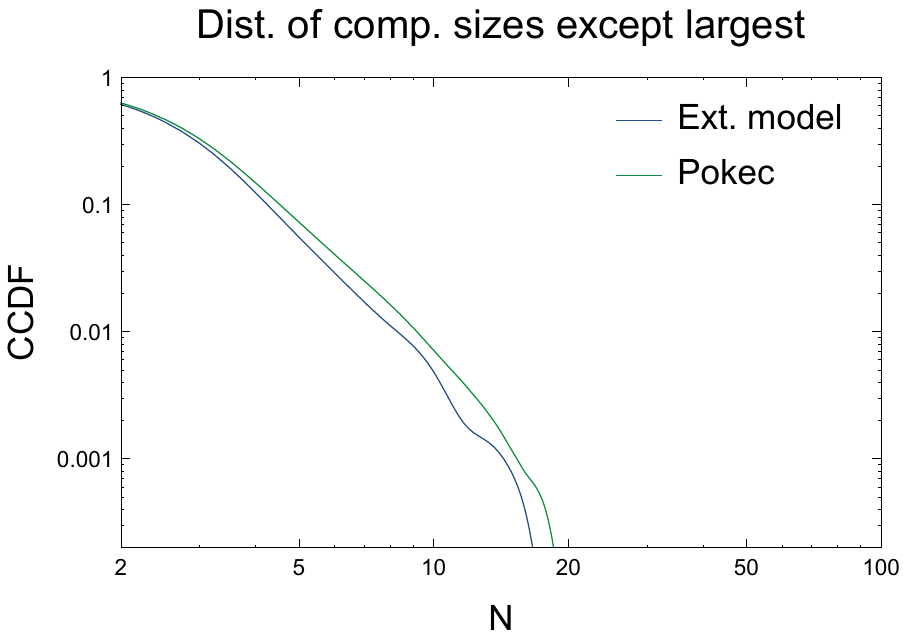}
 \caption{Distribution of component sizes except largest component for the basic 
model (top row) and for the extended model (bottom row). Network sizes are the 
following. \textbf{Left column:} $N=1000$, \textbf{center column:} $N=10000$, 
\textbf{right column:} $N=29000$. The center column shows the distribution of 
sizes near the critical point. One sees that the distribution follows a 
power-law which is expected at the critical point of a phase transition. 
\label{fig_dist_comp}}
\end{figure*}

See Fig. \ref{fig_dist_comp} for distribution of component sizes.

\section{Degree distribution of Pokec OSN}

\begin{figure}[H]
    \centering
 \sifig{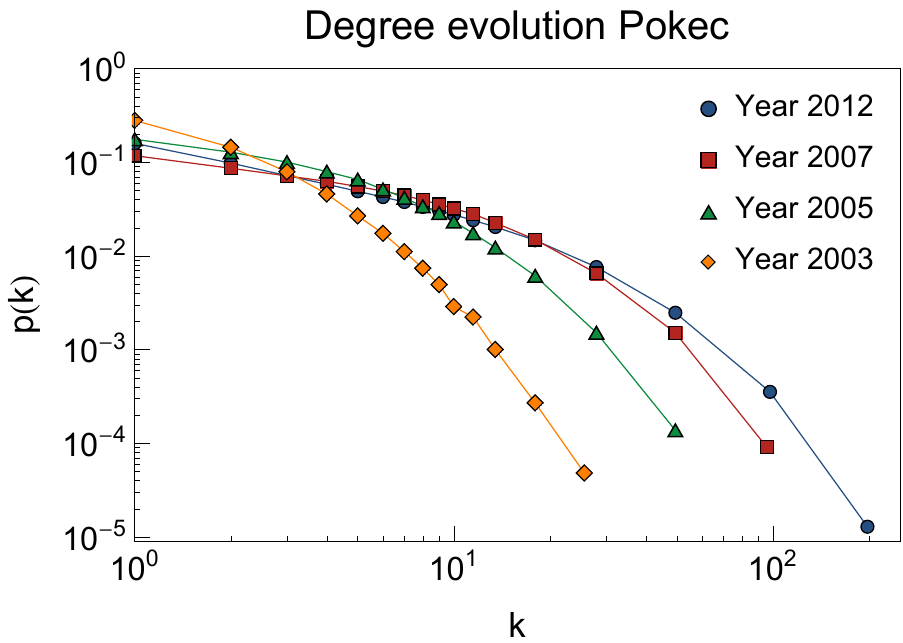}
 \caption{Degree distribution of Pokec OSN for different times. 
\label{fig_pk_pok}}
\end{figure}

\section{Demographics analysis}

 \begin{figure}[H]
  \centering
  \sifig{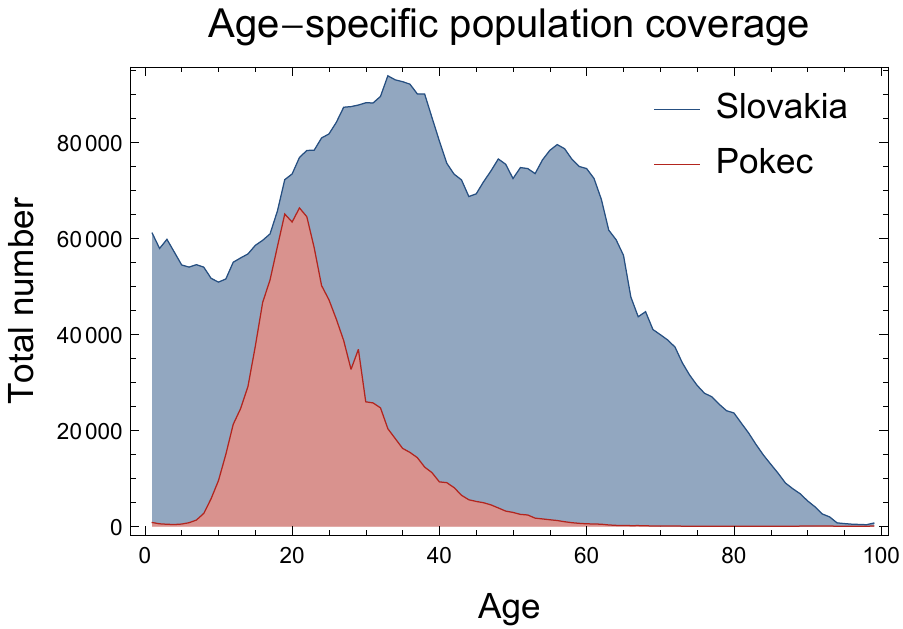}
  \caption{Coverage of Pokec users with respect to the whole population of 
Slovakia. \label{demographic}}
 \end{figure}

\end{document}